\documentclass[12pt,a4paper]{article}
\usepackage{booktabs}
\usepackage{xcolor}
\usepackage{bm}
\usepackage{amsfonts}
\usepackage{amssymb}
\usepackage{mathtools}
\newcommand{\clc}[1]{#1}
\newcommand{\clb}[1]{#1}

\newcommand{\Iv}{{\bf I}}

\newcommand{\vv}{{\bf v}}
\newcommand{\nn}{{\bf n}}
\newcommand{\eperp}{{\bf e}}
\newcommand{\bb}{{\bf b}}
\newcommand{\bg}{{\bf g}}
\newcommand{\tv}{{\bf t}}
\newcommand{\xx}{{\bf x}}
\newcommand{\TT}{{\bf T}}
\newcommand{\xii}{\boldsymbol{\xi}}
\newcommand{\bOmega}{\boldsymbol{\Omega}}
\newcommand{\Div}{\operatorname{div}}
\newcommand{\grad}{\nabla}
\newcommand{\dt}[1]{\dot{#1}}

\newcommand{\Sph}{\mathbb{S}}
\newcommand{\Dtens}{{\bf D}}
\newcommand{\Wtens}{{\bf W}}
\newcommand{\Nvec}{{\bf N}}

\newcommand{\ps}{\partial_s}
\newcommand{\pperp}{\partial_\perp}
\usepackage[left=1.8cm,right=1.8cm,top=2cm,bottom=2cm]{geometry}

\begin{document}

\title{Nonholonomic collective flows: velocity--orientation locking in a continuum with microstructure}

\author{
Gaetano Napoli}

\date{\small Dipartimento di Matematica e Applicazioni "Renato Caccioppoli", Università degli Studi di Napoli "Federico II", Napoli, Italy}




\maketitle

\begin{abstract}
We develop a continuum theory for a fluid of elongated particles that advance along
their own axes. The same kinematics is ideally shared by flocks of sheep,
self-propelled rods, vehicular traffic, and turning flocks of birds. Within the
framework of continua with vectorial microstructure, in which each point also carries
an orientation, we impose the no-side-slip (skate) condition $\vv=u\,\nn$ as an ideal,
non-integrable internal constraint. We derive the pure equations of motion and the
equation governing the constraint reaction from the principle of virtual power. 
Using a constitutive closure provided by the Ericksen--Leslie theory of nematic
liquid crystals, we show how the constraint also shapes collective effects. It turns
parabolic orientational diffusion into hyperbolic orientation--density waves and
forbids any steady simple shear. It fixes the empirical Toner--Tu convective
coefficient to the flow-alignment ratio measured in colloidal rollers, thereby
providing a mechanical foundation for angular sound in micropolar active hydrodynamics. It also
recovers the inertial spin model of bird flocks, with the addition of a banking force
and a turn--density coupling. Finally, it recovers the classical macroscopic models of
one-dimensional traffic flow, while describing steering, lateral tyre forces, and road
geometry in two dimensions.
\end{abstract}



\section{Introduction}
\label{sec:sota}

We start with a simple picture. A flock of sheep can be viewed as a continuum whose
elements are short rods. Each rod tends to move along its own long axis: it may turn,
but it does not slide sideways. The same picture applies to self-propelled rods, cars
on a road, and birds in a turning flock. In all these systems the velocity is tied to
the particle orientation, and motion along the axis is the only motion allowed. Our
aim is to build a continuum theory for such a fluid, and to draw out the consequences
of the sideways constraint for the kinematics, the equations of motion, and the stress.

This kinematics is not just a metaphor. Living systems realize it, at both the
single-rod and the many-body level. The sharpest single-rod example is cytoskeletal
filaments in gliding motility assays --- microtubules propelled by surface-bound
kinesin, or actin filaments by myosin~\cite{Sanchez2012,Schaller2010}. They glide by
following their own contour: the leading tip sets the direction, and the trailing
material follows in turn, each element moving along its axis without slipping
sideways. This is the kinematic signature of a nonholonomic constraint, the
no-side-slip condition of a skater. When densely packed, the same filaments organize
into active nematics, an experimental counterpart of the theory we develop here.

Gliding filamentous cyanobacteria (\emph{Oscillatoria}, \emph{Phormidium}) give an
equally direct example. They slide over a substrate parallel to their long axis at
speeds of order $10\,\mu\mathrm{m\,s^{-1}}$, driven by type-IV pili or slime
secretion~\cite{Kurjahn2024,FilamentousTopo2024}. Because they rest on a surface,
lateral slip is almost entirely suppressed, and the skate constraint is nearly exact.
Dense mats then show buckling, plectoneme formation, and order-to-structure
transitions, collective flock-like features. A caveat sharpens the picture. At the
low Reynolds numbers of microbial locomotion an elongated body has anisotropic drag,
with the transverse resistance roughly twice the longitudinal one. By itself this is
only a soft bias; it becomes a near-rigid constraint when surface contact suppresses
lateral translation. It is therefore gliders, not free swimmers, that embody the
constraint most closely. The nonholonomic constraint is thus the rigid idealization
of a lateral-slip suppression that is nearly total in gliders and merely frictional
in swimmers.

To turn this picture into a theory, we work in a planar setting, within the framework
of continua with vectorial microstructure, in the unified formulation of
Capriz~\cite{Capriz1989}. Besides its placement, each material point carries an order
parameter on the circle $\Sph^1$, the orientation angle $\theta$. The resulting
theory therefore requires, in addition to the classical balance of linear momentum, a
scalar balance for the particle orientation. On this structure we impose the sideways
constraint. It is linear in the velocity but not integrable, and hence nonholonomic.
The model lies at the meeting point of three strands of literature, and, to our
knowledge, the combination considered here does not appear in the existing
literature. We are careful about the scope of this statement: the
velocity--orientation locking $\vv=u\nn$ is not itself new, since closely related
kinematic structures already appear in several theories of active matter. We review
the three strands in turn --- continua with internal constraints, field theories of
self-propelled rods, and nonholonomic mechanics --- and then state the gap that the
present theory fills.

We begin with continua with internal constraints. Capriz and
colleagues~\cite{CaprizPodioGuidugli1984,CaprizPodioGuidugli1983}, together with
related work of the 1980s, studied such constraints in continua with microstructure.
The constraints they analysed are holonomic: restrictions on the configurations a
body element may take, such as rigidity or the normalization $|\nn|=1$ of a
director~\cite{Capriz:2001}. These constraints generate reactive contributions to the
stress and microstress, the fields conjugate to the macroscopic and internal degrees
of freedom. In some cases the constraint even freezes part of the microstructure,
which simplifies the kinematics but complicates the stress, and hence the equations
of motion or equilibrium~\cite{capriz:1985}. Our constraint is of a different kind: it
restricts the velocity, not the configuration, and does not appear to belong to this
class.

The second strand is the physics of self-propelled rods. Several continuum field
theories describe the density and orientation of aligning rods, together with band
formation, active turbulence, and the associated
instabilities~\cite{TonerTu1995,Marchetti2013,Peshkov2012}. In all of them the
alignment between velocity and axis is statistical: it emerges as an ordered state
from steric interactions, friction, and noise. It is neither an exact kinematic
constraint nor the source of a nonholonomic reaction. Closest to our model is
self-organized hydrodynamics (SOH)~\cite{DegondMotsch2008}, the hydrodynamic limit of
the time-continuous Couzin--Vicsek algorithm~\cite{Vicsek1995,Couzin2002}. It couples
a conservation law for the density to a non-conservative, hyperbolic equation for the
director of the mean velocity, with $\vv=c_0\bOmega$ and $|\bOmega|=1$. This is
precisely the slaving $\vv=u\nn$ of the present model, and the later SOH literature
builds on it, with results on well-posedness, on coupling with Navier--Stokes, and on
density-dependent speeds. The difference lies in the mechanical origin of the
condition. In SOH the normalization $|\bOmega|=1$ is imposed by kinematic
construction in the mean-field limit; it is not derived as an ideal nonholonomic
constraint. As a result the transverse reaction $\lambda\,\eperp$, with its
centripetal term $\rho\,u\dot\theta$, is absent, and so is the d'Alembert reading of a
non-integrable constraint with a workless reaction.

The third strand is nonholonomic mechanics. A single such particle is a Chaplygin
sleigh, or knife-edge, the classical example in analytical mechanics: the constraint
$-\dot x\sin\theta+\dot y\cos\theta=0$ is non-integrable~\cite{Bloch2003,NeimarkFufaev1972}.
To our knowledge, such an element has not been homogenized into a continuum with
vectorial microstructure, that is, into a field of coupled Chaplygin sleighs. That
nonholonomic internal constraints with reactions are legitimate objects of study is
confirmed by a different problem: volumetric growth has recently been cast as a
mechanical problem with a nonholonomic, rheonomic constraint~\cite{GrilloDiStefano2023}.

Against these strands, our contribution is threefold. First, we classify the skate
condition as a nonholonomic internal constraint, distinct both from the holonomic
internal constraints of the Capriz tradition and from the constitutive imposition of
SOH. Second, its reaction has a specific structure: a holonomic constraint on
$\grad\vv$ gives a reactive stress, whereas a nonholonomic constraint algebraic in
$\vv$ gives a reactive body force $\lambda\,\eperp$ that cannot be written as
$\Div\TT$. Third, we bridge the single particle and the continuum, obtaining a field
of Chaplygin sleighs coupled through the stress and the microstress. Our novelty
claim is therefore specific. Velocity--orientation locking is not new, and it should
be compared systematically against (i) nonholonomic continuum mechanics with
microstructure, (ii) continuum limits of Chaplygin-sleigh-type or wheeled-particle
systems, (iii) active-matter models with exact velocity--polarization locking, and
(iv) kinematically constrained polar active fluids of Toner--Tu type. The central
object is therefore not velocity--orientation locking by itself, but its mechanical
realization as an ideal nonholonomic constraint in a microstructured continuum,
together with the associated transverse reaction $\lambda\,\eperp$.

The paper is organized as follows. Sections~\ref{sec:kinematics}--\ref{sec:constraint}
set up the kinematics, obtain the balance laws from the principle of virtual power,
impose the skate constraint, and derive its reaction and the reduced equations of
motion. Section~\ref{sec:thermo} verifies thermodynamic consistency, and
Section~\ref{sec:EL} shows that constrained Ericksen--Leslie theory provides a
constitutive closure, while Section~\ref{sec:stress} discusses the active terms.
Section~\ref{sec:benchmark} solves a discriminating benchmark, and
Section~\ref{sec:tonertu} builds the mechanical foundation of Toner--Tu sound.
Sections~\ref{sec:traffic} and~\ref{sec:flocking} give the applications to vehicular
traffic and inertial flocking, respectively, and Section~\ref{sec:outlook} concludes.

\section{Kinematics and balance laws}
\label{sec:kinematics}

\begin{figure}[t]
\centering
\includegraphics[width=\textwidth]{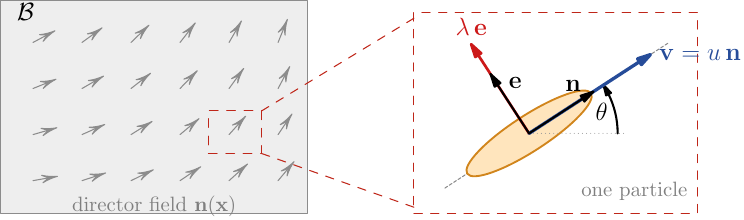}
\caption{Schematic of the model. Left: the body $\mathcal B$ is a continuum of elongated particles; the grey segments are the local director field $\nn(\xx)$. Right: a zoom on a single particle. Its axis defines the unit vector $\nn =(\cos\theta,\sin\theta)$ and the transverse unit vector $\eperp=(-\sin\theta,\cos\theta)$. The skate constraint forbids lateral motion, $\vv\cdot\eperp=0$, so the velocity is parallel to the axis. The transverse reaction $\lambda\,\eperp$ (red) is the force that enforces the constraint.}
\label{fig:schematic}
\end{figure}

We work in two spatial dimensions. Besides its macroscopic placement
$\xx$ and velocity $\vv=\dt{\xx}$, each material point carries an orientation
angle $\theta\in\Sph^1$, with unit axis and transverse unit vector
\begin{equation}
\nn=(\cos\theta,\sin\theta),\qquad
\eperp=(-\sin\theta,\cos\theta),\qquad
\dt{\nn}=\dot\theta\,\eperp .
\end{equation}
The microstructural velocity is the angular rate $\dot\theta$, and $\chi$ denotes
the microinertia (moment of inertia per unit mass). The independent kinematic
fields are, before the constraint, $(\vv,\theta)$ with rates $(\vv,\dot\theta)$.
Figure~\ref{fig:schematic} illustrates the continuum and the notation for a single
particle, together with the skate constraint discussed in
Section~\ref{sec:constraint}.

We now turn to the balance laws. In Capriz's formulation they are postulated. We
obtain them instead, equivalently, from the principle of virtual power. For independent
virtual velocities $(\delta\vv,\delta\dot\theta)$ the internal virtual power is
\begin{equation}
P_{\mathrm{int}}=\int_{\mathcal B}
\Big(\TT \cdot \grad\delta\vv+\xii\cdot\grad\delta\dot\theta+\pi\,\delta\dot\theta\Big)\,\mathrm dV,
\end{equation}
where $\TT$ is the Cauchy stress, $\xii$ the vectorial microstress, that is, an
orientational couple stress conjugate to $\grad\dot\theta$, and $\pi$ the scalar
internal self-force conjugate to $\dot\theta$. With external and inertial powers
\begin{align}
P_{\mathrm{ext}}&=\int_{\mathcal B}(\rho\,\bb\cdot\delta\vv+\rho\beta\,\delta\dot\theta)\,\mathrm dV
+\int_{\partial\mathcal B}(\tv \cdot\delta\vv+s\,\delta\dot\theta)\,\mathrm dA,\\
P_{\mathrm{in}}&=\int_{\mathcal B}(\rho\dt{\vv}\cdot\delta\vv+\rho\chi\ddot\theta\,\delta\dot\theta)\,\mathrm dV,
\end{align}
the balance identity requires $P_{\mathrm{in}}=P_{\mathrm{ext}}-P_{\mathrm{int}}$ for all
admissible virtual velocities $(\delta\vv,\delta\dot\theta)$. Applying the divergence theorem
to integrate by parts the stress power terms, we have
\begin{equation}
\int_{\mathcal B}\TT \cdot \grad\delta\vv\,\mathrm dV = \int_{\partial\mathcal B}(\TT\boldsymbol{\nu})\cdot\delta\vv\,\mathrm dA - \int_{\mathcal B}(\Div\TT)\cdot\delta\vv\,\mathrm dV,
\end{equation}
and similarly for $\xii\cdot\grad\delta\dot\theta$, where $\boldsymbol{\nu}$ is the outward
unit normal to the boundary $\partial\mathcal B$. Substituting these into the power identity
and using the arbitrariness of $\delta\vv$ and $\delta\dot\theta$ in the bulk and on the
boundary, we obtain the two Capriz local balances
\begin{align}
\rho\dt{\vv}&=\Div\TT+\rho\bb, \label{eq:macro}\\
\rho\chi\ddot\theta&=\Div\xii-\pi+\rho\beta, \label{eq:micro}
\end{align}
with boundary conditions $\TT\boldsymbol{\nu}=\tv$ and $\xii\cdot\boldsymbol{\nu}=s$.

Invariance of
$P_{\mathrm{int}}$ under a superposed rigid rotation gives the moment-of-momentum
identity. For a rigid virtual motion of angular velocity $\omega$ one has
$\delta\dot\theta=\omega$ and $(\grad\delta\vv)_{12}=-\omega$,
$(\grad\delta\vv)_{21}=\omega$, so that the internal power reduces to
$P_{\mathrm{int}}=\int_{\mathcal B}\omega\,(T_{21}-T_{12}+\pi)\,\mathrm dV$. Its
vanishing for every $\omega$ gives the identity, in tensorial form,
\begin{equation}
\epsilon_{ijk}\,T_{jk}=\pi\,e_i ,
\label{eq:skew}
\end{equation}
where $\epsilon_{ijk}$ is the three-index permutation (Ricci) symbol and $\bf e$ is
the unit vector normal to the plane, along the microrotation axis; explicitly,
\begin{equation}
T_{12}-T_{21}=\pi .
\label{eq:skew-explicit}
\end{equation}
The stress is therefore in general non-symmetric. Its skew part is carried by the
orientational self-force $\pi$, while $\xii$ acts as a couple stress. This holds
already before we impose the constraint.

\section{The skate constraint and its reaction}
\label{sec:constraint}

A skate advances along its blade but does not slip sideways. The velocity
component transverse to the axis therefore vanishes,
\begin{equation}
C:=\vv\cdot\eperp=-v_x\sin\theta+v_y\cos\theta=0 .
\label{eq:constraint}
\end{equation}
Equation~\eqref{eq:constraint} is linear in the velocity but not integrable. It is
the classical knife-edge, or Chaplygin, constraint, and hence nonholonomic. It
slaves the velocity to
\begin{equation}
\vv=u\,\nn,\qquad u:=\vv\cdot\nn,
\label{eq:slaving}
\end{equation}
so the independent kinematic fields reduce to $(u,\theta)$ (Figure~\ref{fig:schematic}).

When the configuration is frozen, the admissible virtual macro-velocities satisfy
$\delta\vv\cdot\eperp=0$, while $\delta\dot\theta$ remains free (the constraint
does not restrict rotation: a skate turns freely). We take the constraint to be
\emph{ideal}, so its reaction does no virtual power on admissible
velocities. We verify this ideality thermodynamically in
Section~\ref{sec:thermo}. It is not automatic for nonholonomic constraints in
continua~\cite{RajagopalSaccomandi2006}. A multiplier field $\lambda$ then enters
through $\int_{\mathcal B}\lambda\,(\delta\vv\cdot\eperp)\,\mathrm dV$.
Because the constraint is algebraic in the velocity, and not in its gradient, the
reaction is not a stress but a transverse reactive body force,
\begin{subequations}\label{eq:macro-constrained}
\begin{align}
\rho\dt{\vv}&=\Div\TT+\rho\bb+\lambda\,\eperp, \label{eq:macro-c-a}\\
\rho\chi\ddot\theta&=\Div\xii-\pi+\rho\beta . \label{eq:macro-c-b}
\end{align}
\end{subequations}
No reactive microforce conjugate to $\delta\dot\theta$ appears. The nonholonomic
reaction acts on translation only, and orthogonally to the axis. The skate
constraint does not involve the microrotation rate $\dot\theta$, so its ideal
Chetaev (Lagrange--d'Alembert) reaction has no component conjugate to $\dot\theta$.
The constraint therefore reacts on translation alone and leaves the substructural
balance~\eqref{eq:macro-c-b} formally unchanged. A vakonomic reading, in which the
configuration is also varied inside the constraint, would instead couple the
constraint into the director balance. The two prescriptions are inequivalent, and
we adopt the Lagrange--d'Alembert one, appropriate to a genuine mechanical
constraint~\cite{Bloch2003,Lemos2022}.

This is the structural difference from holonomic internal constraints. A holonomic
constraint on the velocity gradient, such as incompressibility $\Div\vv=0$,
restricts $\grad\vv$ and produces a reactive stress $-p\Iv$. The skate
constraint, which is algebraic in $\vv$ and nonholonomic, produces instead a
reactive body force $\lambda\,\eperp$ that is not, in general, the divergence of a
stress field.
Using $\dt{\vv}=\dot u\,\nn+u\dot\theta\,\eperp$ and projecting
\eqref{eq:macro-c-a} onto $\eperp$ gives the following equation
\begin{equation}
\rho\,u\dot\theta=\eperp\cdot\Div\TT+\rho\,\bb\cdot\eperp+\lambda ,
\label{eq:normal}
\end{equation}
whence the reaction is determined,
\begin{equation}
\lambda=\rho\,u\dot\theta-\eperp\cdot\Div\TT-\rho\,\bb\cdot\eperp ,
\label{eq:lambda}
\end{equation}
where the term $\rho\,u\dot\theta$ is a
distributed transverse tension. Wherever the advancing speed and the axis rotation are both present, a lateral reaction must supply the centripetal acceleration that keeps the velocity aligned with the turning axis. This is the continuum analogue of the force exerted by the blade on a skate.

The closed system, given constitutive laws for $\TT,\xii,\pi$, consists of mass conservation and the tangential balance,
\begin{subequations}\label{eq:closed}
\begin{align}
\dot\rho+\rho\,\Div(u\nn)&=0, \label{eq:mass}\\
\rho\dot u&=\nn\cdot\Div\TT+\rho\,\bb\cdot\nn, \label{eq:tang2}
\end{align}
\end{subequations}
together with the substructural balance~\eqref{eq:macro-c-b}, and with $\lambda$
recovered afterwards from \eqref{eq:lambda}. The direction of the
flow is entirely slaved to $\theta$, which evolves through the microstructural
balance. The nonholonomic reaction has no evolution equation of its own, and is determined once the fields are known.

\section{Thermodynamic consistency}
\label{sec:thermo}

We now verify that the theory is compatible with the second law. The skate
constraint is linear and homogeneous in the velocity, so the actual velocity is
itself an admissible virtual velocity. By the assumed ideality, the nonholonomic
reaction $\lambda\,\eperp$ is then workless on the actual motion, and it drops out
of the dissipation inequality. We follow the Coleman--Noll
procedure~\cite{ColemanNoll1963,GurtinFriedAnand2010} to restrict the remaining
constitutive terms. The new features are mechanical, so it is enough to work with
the isothermal mechanical dissipation inequality. Heat conduction enters in the
standard way, and we omit it.

Let $\psi$ be the Helmholtz free energy per unit mass. The isothermal form of the
second law requires that, for every process, the internal working exceed the rate
of stored energy,
\begin{equation}
\mathcal D:=
\TT\cdot\grad\vv+\xii\cdot\grad\dt\theta+\pi\,\dt\theta
+\lambda\,(\vv\cdot\eperp)
-\rho\,\dt\psi\;\ge\;0 .
\label{eq:CD}
\end{equation}
The reaction contributes to \eqref{eq:CD} only through the term
$\lambda\,(\vv\cdot\eperp)$. Because the actual motion satisfies the constraint
$\vv\cdot\eperp=0$ identically, we have
\begin{equation}
\lambda\,(\vv\cdot\eperp)=0 .
\label{eq:reaction-workless}
\end{equation}
As we expected, \emph{the ideal nonholonomic reaction is identically workless and
drops out of the dissipation inequality}. It requires no thermodynamic restriction.
This agrees with the exact energy conservation of the underlying Chaplygin
sleigh~\cite{Bloch2003}. This is a property of the \emph{particular}
skate constraint, whose reaction is orthogonal to the admissible velocity. It is
not a general feature of nonholonomic constraints in continua, which may in
principle be dissipative~\cite{RajagopalSaccomandi2006}.

Take the free energy to depend on density and on the orientation gradient,
$\psi=\psi(\rho,\grad\theta)$, the natural choice for a compressible medium with
Frank elasticity. Then
\begin{equation}
\rho\,\dt\psi=\rho\,\frac{\partial\psi}{\partial\rho}\,\dt\rho
+\rho\,\frac{\partial\psi}{\partial\grad\theta}\cdot\grad\dt\theta
-\rho\,\frac{\partial\psi}{\partial\grad\theta}\cdot(\grad\vv)^{\!\top}\grad\theta,
\end{equation}
where the last group collects the terms by which convection rotates $\grad\theta$.
Using mass balance $\dt\rho=-\rho\,\Div\vv$ and splitting each action into a
reversible (equilibrium) and a dissipative part,
\begin{equation}
\TT=\TT^{\mathrm{eq}}+\TT^{\mathrm{dis}},\qquad
\xii=\xii^{\mathrm{eq}}+\xii^{\mathrm{dis}},\qquad
\pi=\pi^{\mathrm{eq}}+\pi^{\mathrm{dis}},
\end{equation}
the Coleman--Noll argument (arbitrariness of the process rates) identifies the
reversible parts with the derivatives of $\psi$,
\begin{subequations}\label{eq:reversible}
\begin{align}
\TT^{\mathrm{eq}}&=-p\,\Iv-\rho\,\frac{\partial\psi}{\partial\grad\theta}\otimes\grad\theta,
\qquad p=\rho^2\frac{\partial\psi}{\partial\rho}, \label{eq:reversible-T}\\
\xii^{\mathrm{eq}}&=\rho\,\frac{\partial\psi}{\partial\grad\theta}. \label{eq:reversible-xi}
\end{align}
\end{subequations}
Here $-p\Iv$ collects the pressure and $-\rho(\partial\psi/\partial\grad\theta)\otimes\grad\theta$
is the Ericksen elastic (Frank) stress.

What remains of \eqref{eq:CD} is the residual dissipation carried by the
dissipative actions alone,
\begin{equation}
\mathcal D_{\mathrm{dis}}=
\TT^{\mathrm{dis}}\cdot\grad\vv+\xii^{\mathrm{dis}}\cdot\grad\dt\theta
+\pi^{\mathrm{dis}}\,\dt\theta\;\ge\;0 .
\label{eq:residual}
\end{equation}
Objectivity reduces the arguments to the strain rate $\Dtens$ and to the
corotational rate
\begin{equation}
\Nvec=\dt{\nn}-\Wtens\nn=(\dt\theta-w)\,\eperp,
\label{eq:N}
\end{equation}
so that a general linear (Newtonian) closure reads
$\TT^{\mathrm{dis}}=\mathbb L[\Dtens,\Nvec]$ and 
$\pi^{\mathrm{dis}}=\gamma_1(\dt\theta-w)+\gamma_2(\eperp\!\cdot\!\Dtens\nn)$,
with $\mathbb L$ the viscosity tensor and $\gamma_1, \gamma_2$ the rotational 
and flow-alignment viscosities~\cite{Stewart2004}.
Inequality \eqref{eq:residual} then demands positive semi-definiteness of the
dissipation potential\clc{, i.e.\ of the quadratic form built from $\mathbb L$,
$\gamma_1$ and $\gamma_2$; the explicit coefficient restrictions in the
Ericksen--Leslie closure are collected in Section~\ref{sec:EL}.}

Under these conditions, the theory is therefore
thermodynamically consistent in its passive part, with the nonholonomic reaction
contributing nothing to the dissipation.
 
\section{Relation to Ericksen--Leslie theory}
\label{sec:EL}

Ericksen--Leslie (EL) nematodynamics is the standard continuum theory of a fluid
whose microstructure is a director~\cite{Ericksen1961,Leslie1968}. There the
velocity and the director are independent fields. When we impose the skate
constraint $\vv=u\nn$, the two fields are tied together. Then, as we now show,
compressible planar EL becomes a constitutive closure of the nonholonomic continuum of
Sections~\ref{sec:kinematics}--\ref{sec:stress}. Throughout this section an overdot
denotes the material derivative $\tfrac{D}{Dt}=\partial_t+\vv\cdot\grad$, while
$\ps:=\nn\cdot\grad$ and $\pperp:=\eperp\cdot\grad$ are the axial and transverse
directional derivatives.

To bridge the two formulations, we unify the notation by identifying the components
of the Cauchy stress $\TT = \TT^{\mathrm{eq}} + \TT^{\mathrm{dis}}$ derived in Section~\ref{sec:thermo}
with the standard EL tensors. 

In the general linear formulation for a compressible nematic liquid crystal~\cite{Leslie1968},
the dissipative stress $\TT^{\mathrm{dis}}$ is identified with the full Leslie viscous stress $\TT^{v}$.
To ensure material objectivity, $\TT^{v}$ must depend linearly on both the corotational rate
$\Nvec$ defined in \eqref{eq:N} and the symmetric strain rate $\Dtens=\operatorname{sym}\grad\vv$.
In its most general compressible form, this involves nine independent viscosity coefficients:
\begin{align}
\TT^{v}&=\alpha_1(\nn\cdot\Dtens\nn)\nn\otimes\nn + \alpha_2\Nvec\otimes\nn + \alpha_3\nn\otimes\Nvec \nonumber\\
&\quad +\alpha_4\Dtens + \alpha_5(\Dtens\nn)\otimes\nn + \alpha_6\nn\otimes(\Dtens\nn) \nonumber\\
&\quad +\alpha_7(\operatorname{tr}\Dtens)\Iv + \alpha_8(\operatorname{tr}\Dtens)\nn\otimes\nn + \alpha_9(\nn\cdot\Dtens\nn)\Iv,\\
\bg^{v}&=-\gamma_1\Nvec - \gamma_2\Dtens\nn,
\end{align}
where the rotational and flow-alignment viscosities are
\begin{equation}
\gamma_1=\alpha_3-\alpha_2,\qquad \gamma_2=\alpha_6-\alpha_5=\alpha_2+\alpha_3,
\end{equation}
the last identity being the Parodi relation~\cite{Parodi1970}. Because the medium is
compressible, we do not impose $\Div\vv=0$. 

The thermodynamic dissipation inequality requires the residual dissipation to be non-negative
for all independent processes: 
\begin{equation}
\mathcal{D}_{\mathrm{dis}} = \TT^{v}:\Dtens - \bg^{v}\cdot\Nvec \ge 0
\end{equation}
This imposes a strict set of thermodynamic restrictions on the nine coefficients~\cite{Leslie1968,Stewart2004}.
Specifically, for the shear and rotational dynamics, one must have 
\[
\gamma_1 \ge 0, \qquad \alpha_4 \ge 0,
\]
\[
2\alpha_4 + \alpha_5 + \alpha_6 \ge 0, \qquad 4\gamma_1(2\alpha_4+\alpha_5+\alpha_6) \ge (\alpha_2+\alpha_3+\gamma_2)^2.
\]
Furthermore, the presence of the bulk viscosities $\alpha_7, \alpha_8, \alpha_9$ requires the
quadratic form associated with pure volumetric and axial extensions to be positive semi-definite.
In two dimensions, this translates to the additional bounds 
\[
\alpha_4+\alpha_7 \ge 0, \qquad
\alpha_1+\alpha_4+\alpha_5+\alpha_6+\alpha_7+\alpha_8+\alpha_9 \ge 0,
\]
\[
4(\alpha_4+\alpha_7)(\alpha_1+\alpha_4+\alpha_5+\alpha_6+\alpha_7+\alpha_8+\alpha_9) \ge (2\alpha_7+\alpha_8+\alpha_9)^2.
\]

For the reversible part, we assign the Helmholtz free energy per unit mass as the sum
of a purely volumetric term and the conformational elastic energy,
\begin{equation}
\psi(\rho, \grad\theta) = \psi_0(\rho) + \frac{1}{\rho}W_F(\grad\theta), \qquad W_F = \frac{1}{2} K |\grad\theta|^2,
\end{equation}
where $W_F$ is the one-constant Frank energy density per unit volume. Under this choice,
the thermodynamic pressure is $p=\rho^2\partial\psi/\partial\rho$ and the reversible stress
defined in \eqref{eq:reversible-T} evaluates exactly to $\TT^{\mathrm{eq}} = -p\Iv + \TT^{e}$,
with the Ericksen elastic stress given by
\begin{equation}
\TT^{e} = -\rho\,\frac{\partial\psi}{\partial\grad\theta}\otimes\grad\theta = -K\grad\theta\otimes\grad\theta.
\end{equation}
The molecular field is $h=K\nabla^2\theta$, and in two dimensions the scalar
orientational self-force $\pi$ is the projection $\pi = -\eperp\cdot\bg^v$. Thus, the
momentum and director balances, the latter with the microinertia $\chi$ kept, read
\begin{subequations}\label{eq:EL-balances}
\begin{align}
\rho\,\dt{\vv}&=\Div(-p\Iv + \TT^{v} + \TT^{e}), \label{eq:EL-mom}\\
\rho\chi\,\ddot\theta&=K\nabla^2\theta - \gamma_1(\dt\theta-w) - \gamma_2\,(\eperp\!\cdot\!\Dtens\nn). \label{eq:EL-dir}
\end{align}
\end{subequations}
Notice that the dissipative self-force $\pi^{\mathrm{dis}} = \gamma_1(\dt\theta-w) + \gamma_2(\eperp\!\cdot\!\Dtens\nn)$
is the most general linear objective closure, extending the purely rotational relaxation
by coupling it to the shear rate.

A short computation gives the axial and transverse projections of the divergence of
any second-order tensor $\TT$, with components $T_{\alpha\beta}$ in the moving
orthonormal frame $(\nn,\eperp)$. Using $\grad\nn=\eperp\otimes\grad\theta$,
$\grad\eperp=-\nn\otimes\grad\theta$, $\Div\nn=\pperp\theta$ and
$\Div\eperp=-\ps\theta$, one finds
\begin{align}
\nn\cdot\Div\TT&=\ps T_{nn}+\pperp T_{n\perp}
+(T_{nn}-T_{\perp\perp})\,\pperp\theta-(T_{n\perp}+T_{\perp n})\,\ps\theta,
\label{eq:proj-n}\\
\eperp\cdot\Div\TT&=\ps T_{\perp n}+\pperp T_{\perp\perp}
+(T_{nn}-T_{\perp\perp})\,\ps\theta+(T_{n\perp}+T_{\perp n})\,\pperp\theta.
\label{eq:proj-perp}
\end{align}
The extra, texture-induced terms in \eqref{eq:proj-n}--\eqref{eq:proj-perp}
come from the spatial rotation of the moving frame. They are the continuum
signature of the splay and bend of the director field.

The skate constraint $\vv=u\nn$ forces the velocity gradient to be
$\grad\vv=\nn\otimes\grad u+u\,\eperp\otimes\grad\theta$. Hence, in the $(\nn,\eperp)$ frame,
\begin{equation}
a:=D_{nn}=\ps u,\quad
b:=D_{\perp\perp}=u\,\pperp\theta,\quad
c:=D_{n\perp}=\tfrac12(\pperp u+u\,\ps\theta),\quad
w=\tfrac12(u\,\ps\theta-\pperp u),
\label{eq:frame-comps}
\end{equation}
and the compressibility is explicit: $\Div\vv = \operatorname{tr}\Dtens = a+b = \ps u+u\,\pperp\theta \neq 0$.
Writing $m:=\dt\theta-w$ for the director slip rate, the components of the total Cauchy stress
$\TT=-p\Iv+\TT^{v}+\TT^{e}$  are
\begin{equation}
\begin{aligned}
T_{nn}&=-p+(\alpha_1+\alpha_4+\alpha_5+\alpha_6+\alpha_7+\alpha_8+\alpha_9)\,a+(\alpha_7+\alpha_8)\,b-K(\ps\theta)^2,\\
T_{n\perp}&=\alpha_3\,m+(\alpha_4+\alpha_6)\,c-K\,\ps\theta\,\pperp\theta,\\
T_{\perp n}&=\alpha_2\,m+(\alpha_4+\alpha_5)\,c-K\,\ps\theta\,\pperp\theta,\\
T_{\perp\perp}&=-p+(\alpha_7+\alpha_9)\,a+(\alpha_4+\alpha_7)\,b-K(\pperp\theta)^2 .
\end{aligned}
\label{eq:T-comps}
\end{equation}
The skew part of the stress is exactly $T_{n\perp}-T_{\perp n}=\gamma_1\,m+\gamma_2\,c$.
By the moment-of-momentum identity \eqref{eq:skew}, $\pi = T_{12}-T_{21} = T_{n\perp}-T_{\perp n}$.
Thus, the orientational self-force derived from the stress asymmetry matches
the macroscopic EL dissipative torque $\pi=\gamma_1(\dt\theta-w)+\gamma_2\,c$, ensuring full
consistency without forcing any coefficient to vanish. The bulk viscosities $\alpha_7$, $\alpha_8$, and $\alpha_9$ enter only the diagonal components $T_{nn}$ and $T_{\perp\perp}$; they do not alter the skew part of the stress and are decoupled from the director dynamics.

Inserting \eqref{eq:frame-comps}--\eqref{eq:T-comps} into the projection identities
\eqref{eq:proj-n}--\eqref{eq:proj-perp}, and using
$\dt{\vv}=\dt u\,\nn+u\,\dt\theta\,\eperp$, the momentum balance splits into a genuine
evolution equation for the advancing speed and an algebraic equation for the nonholonomic reaction.
At the same time, the director balance \eqref{eq:EL-dir} becomes an evolution for
$\theta$ slaved to the identical kinematic texture,
\begin{equation}
\rho\chi\,\ddot\theta=K\nabla^2\theta
-\gamma_1\Big(\dt\theta-\tfrac12(u\,\ps\theta-\pperp u)\Big)
-\gamma_2\,\tfrac12(\pperp u+u\,\ps\theta)+\rho\beta.
\label{eq:EL-director}
\end{equation}

Because we use the compressible framework, $u$ is a fully dynamical field. It is
governed by mass balance $\dt\rho+\rho\,(\ps u+u\,\pperp\theta)=0$ alongside
\eqref{eq:tang2}, while the transverse body force $\lambda\,\eperp$ simply
accommodates the constraint afterwards through \eqref{eq:lambda}.

\section{\clc{Structure of} the active terms}
\label{sec:stress}

We model the tendency of the particles to advance along their axis as an active
self-propulsion. In the literature, this enters either as an active
stress
\begin{equation}
\TT_a=\sigma_a\,\nn\otimes\nn
\end{equation}
or as an external body force
\begin{equation}
\rho\bb=\rho f_0\,\nn,
\end{equation}
both of which feed the tangential momentum equation~\eqref{eq:tang2}.
Self-propulsion and active stress need not satisfy the dissipation
inequality~\eqref{eq:residual}. They are a sustained conversion of internal
(metabolic or fuel) energy into mechanical work, so the medium is a non-equilibrium
active system~\cite{Marchetti2013}. For this reason we follow
\clc{Turzi}~\cite{Turzi2017} and describe self-propulsion by an external active
force that competes with the passive relaxation. This is the clearest and most
consistent route thermodynamically, and it avoids ambiguities with the internal
stress tensor. The single element is then a driven Chaplygin sleigh, and the
continuum is a field of coupled sleighs bound together by $\TT$, $\xii$, and the
transverse tension $\lambda\,\eperp$.

Besides propulsion, active systems admit an \emph{active reorientation}. It is an
external microforce that turns the particle axis along transverse density gradients,
\begin{equation}
\rho\,\beta_{\mathrm a}=-\kappa\,\eperp\!\cdot\!\grad\rho ,
\label{eq:active-torque}
\end{equation}
where $\kappa$ is a constitutive coefficient. It is the continuum form of the
density-dependent alignment used in flocking
models~\cite{DegondMotsch2008,TonerTu1995}. It enters the substructural
balance~\eqref{eq:macro-c-b}, where it couples the orientation to the density. It is
the exact ingredient that produces the orientation--density waves of
Section~\ref{sec:benchmark}.

Unlike tangential self-propulsion, the orientational coupling~\eqref{eq:active-torque}
could in principle be read either reversibly or as a non-equilibrium term.
In polar fluids, such a term could arise reversibly from a free energy that couples
the director $\nn$ directly to the density gradient $\grad\rho$. But a density
gradient in the energy potential would turn the medium into a
Korteweg fluid (as seen, for instance, \clc{in} Ref.~\cite{selinger:2002}), and
this would then require a structural change of the reversible Cauchy
stress. In the macroscopic limits of flocking problems~\cite{DegondMotsch2008}, by
contrast, this term is treated as a non-equilibrium input that reflects the agents'
sensory responses.

Beyond this choice, the form of the torque follows from the structure of the
theory. The substructural balance~\eqref{eq:macro-c-b} is an angular-momentum balance
for the axis $\nn$, so an external body couple $\rho\beta_{\mathrm a}$ must be a scalar
that rotates $\nn$. At leading order the only field the local environment supplies is
the density gradient $\grad\rho$. The unique objective scalar linear in it that
rotates the axis is the transverse projection $\eperp\!\cdot\!\grad\rho$.
Hence $\rho\beta_{\mathrm a}=-\kappa\,\eperp\!\cdot\!\grad\rho$ is the lowest-order
admissible steering response, not an arbitrary import. Its meaning is direct. For
$\kappa>0$ the axis turns down the density gradient, away from congestion. This is
the continuum form of a driver leaving a crowded lane, or of an agent avoiding a
dense region. The nonholonomic constraint $\vv=u\nn$ then turns this reorientation
into a genuine transverse motion, so the same $\kappa$ sets the speed of the
orientation--density waves. A first-principles derivation of $\kappa$ from a
density-gradient energy (a Korteweg term) or from a kinetic model remains open. Here
we keep it as the leading phenomenological coupling, in line with the active-matter
literature.



\section{A discriminating benchmark}
\label{sec:benchmark}

We now present a problem that can be solved in closed form and that isolates the
effect of the skate constraint. We study the linear response of the aligned state
$\nn_0=\mathbf e_x$, with density $\rho_0$, to a small transverse bend
$\theta_1=\theta_1(y,t)$. In the active case the base state also advances at speed
$u_0$. \clb{Throughout the linear analyses of this and the following two sections we
write $\rho_1,\ \theta_1,\ u_1$ and $v_{1y}$ for the small first-order perturbations
of the density, the orientation, the advancing speed and the transverse velocity
about the base state, reserving the symbol $\delta$ for the virtual variations of
Section~\ref{sec:kinematics}.} The perturbation depends only on the coordinate $y$
transverse to the axis, the transverse geometry~\cite{SimhaRamaswamy2002}. The role
of the constraint rests on one kinematic fact,
\begin{equation}
\vv=u\,\nn\ \Longrightarrow\ v_{1y}=u_0\,\theta_1 .
\label{eq:bench-key}
\end{equation}
Rotating the director produces a transverse velocity, and this pumps mass. The
independent fields of classical Ericksen--Leslie do not do this. To separate
the constraint from the activity, we compare two theories in a passive and in an
active regime: classical EL, where $\vv$ and $\nn$ are independent, and the
skate-constrained theory, where $\vv=u\nn$.

\emph{Passive regime ($u_0=0$).} We keep the fluid compressible throughout, since
the sound is a density phenomenon. Here the coupling $a:=\rho_0 u_0$ vanishes, and
the constraint is silent at linear order. The only motion it forbids is transverse,
whereas the passive response is a longitudinal backflow, which the constraint
already allows. The density fluctuation is still present, but with $a=0$ it no
longer couples to the director. It carries only the ordinary compressible sound.
Both theories then reduce to the same parabolic director relaxation,
\begin{equation}
\gamma_1^{\rm eff}\,\partial_t\theta_1=K\,\partial_y^2\theta_1 ,
\label{eq:bench-EL}
\end{equation}
which implies
\begin{equation}
\omega=-\,i\,\frac{K}{\gamma_1^{\rm eff}}\,q^2 ,
\label{eq:bench-EL-omega}
\end{equation}
with $\gamma_1^{\rm eff}=\gamma_1-\frac{\alpha_2^2}{\eta_b}$, where $\eta_b=\frac{1}{2}(\alpha_3+\alpha_4+\alpha_6)$ is the pertinent Miesowicz shear viscosity.

\emph{Active regime ($u_0\neq0$).} We now fix the activity and compare the two
theories, so that any difference is due to the constraint alone.

In \emph{classical EL} the fields $\vv$ and $\nn$ are independent. We keep the
fluid compressible, since the sound is a density phenomenon. Mass conservation then
couples the density to the transverse velocity, $\partial_t\rho_1+\rho_0\partial_yv_{1y}=0$.
But $v_{1y}$ is an independent field governed by the momentum balance, and at
linear order a bend $\theta_1(y)$ produces no transverse active force. Instead, the
active stress $\sigma_a\,\nn\otimes\nn$ drives a longitudinal shear flow, and this
flow torques the director back through flow alignment. The linearized mass and
director balances thus decouple into
\begin{align}
\partial_t\rho_1+\rho_0\,\partial_yv_{1y}&=0,\\
\gamma_1^{\rm eff}\,\partial_t\theta_1&=K\partial_y^2\theta_1+\nu\,\sigma_a\,\theta_1,
\end{align}
where $\nu\propto\gamma_2$ is the flow-alignment coupling. Since the unforced transverse
velocity $v_{1y}$ rapidly relaxes, the density perturbation is static. Seeking
$\theta_1\propto e^{i(qy-\omega t)}$, the director dispersion relation is
\begin{equation}
\omega = i\left( \frac{\nu\sigma_a}{\gamma_1^{\rm eff}} - \frac{K}{\gamma_1^{\rm eff}}q^2 \right).
\label{eq:bench-classical-active}
\end{equation}
This gives a purely imaginary root: an unconditional growth ($\operatorname{Im}\omega>0$
for long wavelengths, the generic instability of active nematics~\cite{SimhaRamaswamy2002}),
with no propagation ($\operatorname{Re}\omega=0$) and no orientation--density sound.
By contrast, the skate constraint forces $v_{1y}=u_0\theta_1$, replacing
\eqref{eq:bench-classical-active} with the hyperbolic propagating waves of
Eq.~\eqref{eq:bench-disp}.

In the \emph{skate-constrained} theory $\vv=u\nn$, so a bend pumps mass
through~\eqref{eq:bench-key}. Two facts close the reduction and justify a fixed
advancing speed. First, in this transverse geometry the pressure gradient is
orthogonal to the axis, $\nn\cdot\grad p=\partial_x p=0$, so it does not accelerate
the advancing speed. It is carried instead by the transverse reaction
$\lambda\,\eperp$ of Eq.~\eqref{eq:lambda}. The tangential balance~\eqref{eq:tang2}
then reduces to a decoupled, damped equation for $u_1$, which in the strong
speed-relaxation limit is enslaved, $u_1=O(\tau)\to0$, so that $u\equiv u_0$
at leading order. Second, the active reorientation torque~\eqref{eq:active-torque}
supplies the coupling of the director to the density. The mass balance,
compressible so that the density is a dynamical field, and the substructural
balance~\eqref{eq:micro}, in the overdamped limit $\chi\to0$ with
$\pi=\gamma_1(\dt\theta-w)$, give
\begin{align}
\partial_t\rho_1+a\,\partial_y\theta_1&=0, \qquad \qquad (\text{with}\quad a:=\rho_0 u_0)&,
\label{eq:bench-mass}\\
\gamma_1\,\partial_t\theta_1&=K\,\partial_y^2\theta_1-\kappa\,\partial_y\rho_1,
\label{eq:bench-dir}
\end{align}
where the term $-\kappa\,\partial_y\rho_1$ is the linearization of the
active reorientation~\eqref{eq:active-torque}, since
$\eperp\cdot\grad\rho=\partial_y\rho_1$ in this geometry. Seeking
$\rho_1,\theta_1\propto e^{i(qy-\omega t)}$ yields the dispersion relation
\begin{equation}
\gamma_1\,\omega^2+iKq^2\,\omega-\kappa\, a\,q^2=0 ,
\label{eq:bench-disp}
\end{equation}
whence
\begin{equation}
\omega=\frac{-iKq^2\pm\sqrt{4\gamma_1\kappa\, a\,q^2-K^2q^4}}{2\gamma_1}.
\label{eq:bench-omega}
\end{equation}
At long wavelength,
\begin{equation}
\omega\simeq \pm\,c\,q-\frac{iK}{2\gamma_1}\,q^2,
\label{eq:bench-longwave}
\end{equation}
with the speed
\begin{equation}
c=\sqrt{\frac{\kappa\, a}{\gamma_1}}=\sqrt{\frac{\kappa\,\rho_0 u_0}{\gamma_1}} .
\label{eq:bench-speed}
\end{equation}
For $\kappa\, a>0$ these are propagating orientation--density waves with speed
$c$ and diffusive damping $Kq^2/2\gamma_1$; for $\kappa\, a<0$ one root
acquires $\operatorname{Im}\omega>0$, a long-wavelength banding instability. The
governing operator is \emph{hyperbolic}.

The result is robust to the director closure. If we replace the relaxational
coefficient $\gamma_1$ in~\eqref{eq:bench-dir} by the inertial coefficient $\rho_0$
of the SOH direction equation~\cite{DegondMotsch2008}, then \eqref{eq:bench-disp} is
unchanged in form, and hence still hyperbolic. What matters is the coupling
$a=\rho_0u_0\neq0$, which exists only when the constituents self-advance.

The comparison is now clear. In the passive regime the constraint is silent and both theories
diffuse. In the active regime, at the same activity, the only difference is the constraint, and
it replaces the non-propagating generic instability of classical active nematics by
hyperbolic orientation--density waves. This is a change of PDE type, from parabolic
to hyperbolic, produced by the constraint. The mechanism is the loop closed
by~\eqref{eq:bench-key}: a bend $\partial_y\theta_1$ drives the transverse flow
$v_{1y}=u_0\theta_1$, which accumulates density,
$\partial_t\rho_1=-a\,\partial_y\theta_1$; the density gradient torques the
director back through the active reorientation~\eqref{eq:active-torque},
$\gamma_1\partial_t\theta_1\sim-\kappa\,\partial_y\rho_1$; and this restoring
coupling propagates like sound. It is a second orientational sound. It is absent when the
two sectors do not communicate, and it recovers the Toner--Tu flocking
sound~\cite{TonerTu1995}\clc{\cite{TonerTu1998}} as a direct consequence of the constraint. The dispersion
is strongly anisotropic, because the coupling lives only in gradients transverse to
the axis.

The three coefficients of \eqref{eq:bench-mass}--\eqref{eq:bench-dir} have a
transparent meaning. The coupling $a=\rho_0 u_0$ ties mass to orientation, and is
nonzero only under self-advection. The coefficient $\kappa$ of
Eq.~\eqref{eq:active-torque} measures the strength of the active reorientation. The
coefficient $\gamma_1$, or $\rho_0$ in the inertial closure, is the director-response
coefficient, and $K$ is the Frank constant. 
The sign of $\kappa$ decides between waves, for $\kappa>0$, and banding, for
$\kappa<0$.

A second, sharper contrast holds in the passive regime, where the classical
rheological experiment has no analogue.

The skate-constrained nematic admits no steady simple-shear state
$\vv=\dot\gamma\,y\,\mathbf e_x$ with uniform director, except in the degenerate
case $\gamma_1+\gamma_2=0$. Indeed, where $\dot\gamma\,y\neq0$ the velocity points
along $\mathbf e_x$, so the constraint $\vv\parallel\nn$ forces $\nn=\mathbf e_x$,
that is $\theta\equiv0$ and $\dt\theta=0$. For this frozen director the shear gives
$\Dtens_{n\perp}=c=\dot\gamma/2$ and vorticity $w=-\dot\gamma/2$, hence
$N_\perp=\dt\theta-w=\dot\gamma/2$, and the transverse component of the director
balance reads
\begin{equation}
0=K\nabla^2\theta-\gamma_1 N_\perp-\gamma_2 c
=-\tfrac12(\gamma_1+\gamma_2)\,\dot\gamma ,
\end{equation}
since $\nabla^2\theta=0$. This vanishes only for $\dot\gamma=0$ (or
$\gamma_1+\gamma_2=0$), so no non-trivial steady simple shear exists.

The physical reason is simple. The flow-alignment torque tends to rotate the
director, but the constraint drags the velocity with it, so a steady rectilinear
shear cannot persist. A fluid of skates does not let itself be combed by a
transverse shear. Instead it converts the shear into reorientation. The very
response that defines the Leslie viscosities and reveals tumbling~\cite{Stewart2004}
is therefore removed by the constraint, a qualitative mark of the nonholonomic
kinematics.

\section{Mechanical foundation of the Toner--Tu sound}
\label{sec:tonertu}

The orientation--density waves under the skate constraint give a rigorous mechanical
foundation for the flocking sound predicted by the phenomenological
Toner--Tu theory~\cite{TonerTu1995,Marchetti2013}\clc{\cite{TonerTu1998}}. To explore this connection
and compare it with experiment, we extend the analysis to the full
two-dimensional angular sound.

Away from the purely transverse direction, the perturbation also probes the longitudinal
sector. We linearize the reduced system in two dimensions about the flowing state,
keeping the wave vector $\bm q=q(\cos\vartheta,\sin\vartheta)$ at an arbitrary angle
$\vartheta$ to the mean flow. With $\vv=u\nn$ and the advancing speed enslaved in the
strong-relaxation limit ($u\equiv u_0$), mass conservation and the director balance yield
\begin{subequations}\label{eq:sound2d}
\begin{align}
\partial_t\rho_1+u_0\partial_x\rho_1+\rho_0\partial_y v_{1y}&=0,\\
\partial_t v_{1y}+\lambda_1 u_0\partial_x v_{1y}+\sigma\partial_y\rho_1&=D_\perp\partial_y^2 v_{1y} ,
\end{align}
\end{subequations}
where $v_{1y}=u_0\theta_1$ is the transverse velocity. The coefficients are read directly
off the constrained constitutive law:
\begin{equation}
\sigma=\frac{\kappa u_0}{\gamma_1},\qquad D_\perp=\frac{K}{\gamma_1},\qquad
\lambda_1=\tfrac12\Big(1+\frac{\gamma_2}{\gamma_1}\Big).
\label{eq:sound-coeffs}
\end{equation}
The plane-wave analysis then yields the full angular sound speeds
\begin{equation}
2c_\pm(\vartheta)=(1+\lambda_1)\,u_0\cos\vartheta\pm
\sqrt{(1-\lambda_1)^2 u_0^2\cos^2\vartheta+4\sigma\rho_0\sin^2\vartheta}.
\label{eq:angular-sound}
\end{equation}
This is the dispersion form measured in colloidal-roller fluids by Geyer
\emph{et al.}~\cite{Geyer2018}. At $\vartheta=\pi/2$ it reduces to the transverse
sound $c=\sqrt{\sigma\rho_0}$, and the damping to $D_\perp q^2/2$, both reproducing
the reported data. The identification $\sigma=\kappa u_0/\gamma_1$ also clarifies the
apparent dependence on $u_0$. The transverse sound speed is $c=\sqrt{\sigma\rho_0}$,
set by the active compressibility and the density, so sound is faster in denser
fluids, as observed.

The new content lies in the convective coefficient $\lambda_1$. In the
phenomenological Toner--Tu theory it is a free constant, equal to one only for a purely
momentum-conserving fluid. The nonholonomic skate constraint fixes it explicitly,
$\lambda_1=\tfrac12(1+\gamma_2/\gamma_1)$. This ties the broken-Galilean advection of
the sound to the flow-alignment parameter $\gamma_2/\gamma_1$ of the constituents.
It is exactly the same combination $\gamma_1+\gamma_2$ that forbids a steady simple shear.

Two experimental facts support this mechanical identification. First, $\lambda_1$
is a ratio of viscosities, so it is inherently independent of the density, exactly
as reported in the experiments~\cite{Geyer2018}. Second, the measured value
$\lambda_1=0.75\pm0.1$~\cite{Geyer2018} implies, through our relation, an
effective flow-alignment ratio $\gamma_2/\gamma_1=2\lambda_1-1=0.50\pm0.20$.
Because it lies below unity, this ratio places the rollers in the tumbling regime
($|\gamma_2/\gamma_1|<1$) rather than the strongly flow-aligning one. This is
physically natural for the near-spherical Quincke rollers. The Leslie flow-alignment
is set by the particle shape, and it is weak for weakly anisotropic bodies, while
its positive sign reflects the fore--aft polarity of the rolling motion. The value
$\gamma_2/\gamma_1=0.50\pm0.20$ is therefore quantitatively plausible, and it gives
a precise, falsifiable prediction that a direct rheological measurement can test. The
kinetic theory of the same experiment~\clc{\cite{Bricard2013}} is itself built on the
skate assumption, that is, the roller velocity slaved to its orientation, and it
predicts $\lambda_1$ in agreement with the data. Here we recover that agreement not
as a statistical kinetic limit, but as a macroscopic continuum consequence of the
nonholonomic constraint. We express it as a pure constitutive property of the active
fluid.

\section{Application to vehicular traffic}
\label{sec:traffic}

Among the possible realizations of the skate constraint, vehicular traffic is the
most literal\clc{~\cite{Helbing2001}}, because a wheeled vehicle is a nonholonomic
system. Its wheels forbid lateral translation. The car advances along its axis and
steers, but it does not slip sideways. So $\vv=u\nn$ is not an idealization but the
exact rolling constraint, and the single vehicle is a Chaplygin sleigh. We show that
the present continuum contains the classical one-dimensional traffic models as
reductions. We show that it explains their anisotropy as a structural feature, and
that it extends them to a two-dimensional theory with a clear mechanical meaning.

On a straight single-lane road, $\theta\equiv0$, $\nn=\mathbf e_x$, $s=x$, the
director is frozen and the theory reduces to
\begin{subequations}\label{eq:traffic-1d}
\begin{align}
\partial_t\rho+\partial_x(\rho u)&=0, \label{eq:traffic-1d-mass}\\
\rho(\partial_t u+u\,\partial_x u)&=\partial_x T_{xx}+\rho\,b_x . \label{eq:traffic-1d-a}
\end{align}
\end{subequations}
With longitudinal stress $T_{xx}=-P(\rho)+\mu\,\partial_x u$ (a density-dependent
``pressure'' $P$, $P'>0$, plus a viscosity) and a relaxation body force
$b_x=(V(\rho)-u)/\tau$ toward the equilibrium speed--density law $V(\rho)$, this
becomes
\begin{equation}
\rho(\partial_t u+u\,\partial_x u)=-p'(\rho)\,\partial_x\rho+\mu\,\partial_{xx}u
+\rho\,\frac{V(\rho)-u}{\tau}.
\label{eq:traffic-1d-mom}
\end{equation}
The relation to the standard models is summarized in Table~\ref{tab:traffic}.

\begin{table}[htbp]
\centering
\footnotesize
\begin{tabular}{@{} l p{5cm} p{4.5cm} @{}}
\toprule
Model & Governing equations & Character\\
\midrule
LWR & $\rho_t+(\rho V(\rho))_x=0$ & 1st order, $u=V(\rho)$\\[4pt]
Payne--Whitham & $\rho_t+(\rho u)_x=0$ \newline $u_t+u u_x=\frac{V-u}{\tau}-\frac{c_0^2}{\rho}\rho_x$ & 2nd order, isotropic\\[6pt]
Aw--Rascle--Zhang & $\rho_t+(\rho u)_x=0$ \newline $(u{+}p)_t+u\,(u{+}p)_x=\frac{V-u}{\tau}$ & 2nd order, anisotropic\\[6pt]
Present (1D) & Eqs.~\eqref{eq:traffic-1d}--\eqref{eq:traffic-1d-mom} & 2nd order, structural anisotropy\\[4pt]
Present (2D) & Adds Eqs. \eqref{eq:macro-c-b} and \eqref{eq:lambda} & Nonholonomic, coupled yaw dynamics, lateral $g$-force\\
\bottomrule
\end{tabular}
\caption{One-dimensional reduction of the nonholonomic continuum against the
classical traffic models, and its two-dimensional extension.}
\label{tab:traffic}
\end{table}

It has been objected~\cite{Daganzo1995} that second-order models with an isotropic
pressure, such as the Payne--Whitham model~\cite{Payne1971,Whitham1974}, let
disturbances travel faster than traffic, so that vehicles would react to what is
\emph{behind} them. This was resolved~\cite{AwRascle2000,Zhang2002} by making the
pressure \emph{convective}: the Riemann invariant $w=u+p(\rho)$ is advected with
the vehicle, a Lagrangian property of the driver.

Here the anisotropy is automatic, for two reasons. First, in two dimensions the
momentum balance is projected onto $\nn$, the direction of travel, and the
transverse coupling is carried away by the constraint reaction, so that a driver
responds only along the heading. Second, the ARZ convected invariant corresponds
to a convected microstructural order parameter. If we identify the hesitation $p$
with a scalar internal field transported along $\nn$, so that $Dp/Dt$ is a
prescribed source, then the pair $(u,p)$ obeys exactly the ARZ structure. In this
sense ARZ is the one-dimensional, steering-free instance of the nonholonomic
continuum with one convected internal variable, while
LWR~\cite{LighthillWhitham1955,Richards1956} is the further quasi-steady limit
$u=V(\rho)$.

When the heading varies, the reduced system, that is, mass
conservation~\eqref{eq:mass}, the tangential balance~\eqref{eq:tang2}, the
substructural balance~\eqref{eq:macro-c-b}, and the reaction
\eqref{eq:lambda},
acquires a direct vehicular reading. Three identifications are significant.

\emph{(i) Steering is the microstructural dynamics.} The
balance~\eqref{eq:macro-c-b} governs the heading, or yaw, of the flow, and it gives
a mechanical setting for driver behavioral rules. The microstress
$\xii=K\grad\theta$ acts as an alignment stiffness. It models the tendency of a
driver to adjust their heading to match neighboring vehicles (lane discipline).
The active body couple $\rho\beta$ encodes navigational intent. Road tracking
becomes a relaxation $\rho\beta\propto-(\theta-\theta_{\text{road}})$,
while a lane change to escape a congested lane becomes the active
reorientation torque of Eq.~\eqref{eq:active-torque},
$\rho\beta_{\mathrm a}=-\kappa\,\eperp\!\cdot\!\grad\rho$. Finally, $\pi$ gives
the yaw damping (steering viscosity) that models finite reaction times. In this
framework, lane changing and turning become genuine physical balances, not
a phenomenological lateral advection.

\emph{(ii) The reaction is the lateral tyre force.} The term $\rho\,u\,\dt\theta$
is the density of centripetal acceleration, that is, the lateral force the tyres
must supply in a manoeuvre. Thus $\lambda$ predicts the transverse force, or
lateral $g$, a quantity of safety, comfort, and road-holding that the standard
models do not describe.

\emph{(iii) Road geometry is director bend.} Curved roads, roundabouts, and ramps
correspond to a prescribed texture $\nn(\xx)$. Vehicles follow its integral lines,
the autoparallels of the director frame, so that road curvature is encoded in the
bend and torsion of the zweibein. This connects traffic on a network with
transport in the effective geometry of a director
field~\cite{FumeronBerche2022,KatanaevVolovich1992}.

Defects of the heading field organize the flow. Because the velocity is slaved to
the orientation, $\vv=u\nn$, a defect, a point of nonzero winding number,
forces the advancing speed to vanish, $u\to0$. Then $\vv$ stays regular while
$\nn$ is undefined at the core. Defects are therefore stagnation points: a $+1$
vortex is the centre of a roundabout, a source or a sink is an on- or off-ramp
(which enters the mass balance~\eqref{eq:mass} as a source), and a saddle is an
intersection. Near a defect the flow curves sharply, so the transverse reaction
$\lambda=\rho u\dot\theta$, the lateral tyre force, is largest there. Defects thus
mark points that are at once stagnation points and loci of maximal lateral force,
a feature absent from scalar traffic models.

Existing two-dimensional and multilane macroscopic
models~\cite{Multilane2025,TwoDMulticlass2020} add the lateral dynamics through
lateral viscosity, velocity differentials, or quasi-gas-dynamic terms, but they do
not impose the no-side-slip constraint. Macroscopic pedestrian and crowd
models~\cite{Hughes2002,DegondMotsch2008} do use $\vv=u\nn$, but they fix the
direction by an eikonal or a gradient, and their agents are holonomic, since
pedestrians can step sideways. The present model is the nonholonomic, vehicular
counterpart. It keeps the slaving $\vv=u\nn$, but it adds an autonomous heading
balance and the transverse reaction $\lambda$ that holonomic crowd models do not
have. It thus fills, on the traffic side, the same gap it fills for active matter.

\subsection{Numerical demonstration}
\label{sec:traffic-numerical}

To see the macroscopic consequences of the skate constraint in two dimensions, we
simulate a uniform traffic flow that advances at speed $u_0$ and density
$\rho_0$ and encounters a weak, localized obstacle (a slight lane narrowing, or a
slow-moving vehicle). The obstacle acts as a local source of
perturbation $S(x,y)$ in the mass balance. We compare the classical isotropic 2D
extension of macroscopic traffic models with our nonholonomic continuum. The
treatment is deliberately a linear response about the flowing state. It captures the
change of equation type, but it assumes small perturbations. So it does not resolve
the fully nonlinear dynamics near a complete stop, where the advancing speed $u$ itself
would vary strongly and the small-angle relation $v_{1y}\approx u_0\theta_1$ would fail.

In \emph{classical holonomic models}, lateral manoeuvres are often treated by adding
an effective lateral diffusion or an isotropic pressure gradient to the 1D equations
\eqref{eq:traffic-1d-mass}--\eqref{eq:traffic-1d-a}. In the kinematic limit, where
the longitudinal speed relaxes rapidly to $u_0$, the density perturbation $\rho_1$
is simply advected downstream and diffuses laterally. It obeys the parabolic
advection--diffusion equation
\begin{equation}
\partial_t \rho_1 + u_0\,\partial_x \rho_1 = D \nabla^2 \rho_1 + S(x,y),
\label{eq:num_classical}
\end{equation}
where $D$ is the effective macroscopic diffusivity of the crowd of vehicles.

In our \emph{nonholonomic continuum}, the kinematics is fundamentally different.
The skate constraint $\vv = u\nn$ couples the lateral velocity to the
steering angle, so at linear order $v_{1y} \approx u_0 \theta_1$. In the substructural balance
\eqref{eq:macro-c-b}, the drivers' intent to change lane and avoid the congestion
is modeled by the active reorientation torque $\rho\beta_{\mathrm a} = -\kappa\,\partial_y \rho_1$.
We linearize the two-dimensional mass balance \eqref{eq:mass} and the overdamped
heading equation \eqref{eq:macro-c-b} around the base flow. The coupled evolution for
the density and the lateral velocity $v_{1y}$ reads:
\begin{subequations}\label{eq:num_nonholonomic}
\begin{align}
\partial_t \rho_1 + u_0\,\partial_x \rho_1 + \rho_0\,\partial_y v_{1y} &= S(x,y), \\
\partial_t v_{1y} + u_0\,\partial_x v_{1y} + c^2\,\partial_y \rho_1 &= D_\perp \partial_y^2 v_{1y},
\end{align}
\end{subequations}
where $c = \sqrt{\kappa u_0 \rho_0 / \gamma_1}$ emerges as a transverse kinematic wave speed, 
and $D_\perp = K/\gamma_1$ is the orientational diffusion. The system
\eqref{eq:num_nonholonomic} is structurally \emph{hyperbolic}.

\begin{figure}[htbp]
    \centering
    \includegraphics[width=\textwidth]{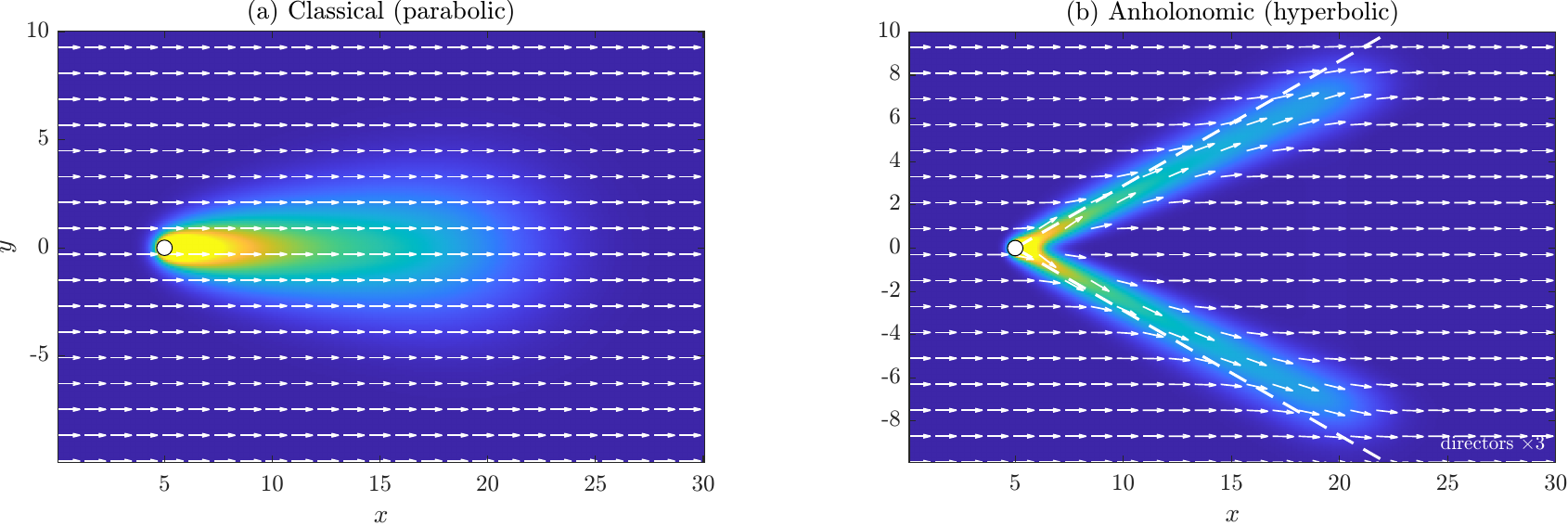}
    \caption{Stationary response to a weak localized obstacle (white dot) in a uniform traffic flow advancing from left to right at speed $u_0$. \textbf{(a)} In the classical parabolic formulation \eqref{eq:num_classical}, the avoidance manoeuvre produces a diffusive boundary layer (wake) that widens downstream. \textbf{(b)} In the nonholonomic continuum \eqref{eq:num_nonholonomic}, the skate constraint forces the velocity to follow the perturbed heading. The system is hyperbolic, and the lateral evasion propagates transversally at the kinematic sound speed $c$, forming a distinct Mach cone. Dashed lines mark the theoretical Mach angle $\sin\alpha = c/u_0$.}
    \label{fig:mach_cone}
\end{figure}

We solve Eqs.~\eqref{eq:num_classical} and \eqref{eq:num_nonholonomic} numerically using 
a finite-difference time-domain method with upwind advection and open boundaries. 
The stationary states are shown in Figure~\ref{fig:mach_cone}.

The two responses differ sharply. In the classical case (Figure~\ref{fig:mach_cone}a),
the obstacle produces a smooth, parabolic wake. The lateral extent of the 
perturbed region grows diffusively downstream as $y \sim \sqrt{Dx/u_0}$. The flow 
simply seeps around the obstacle without a well-defined shock front.

In the nonholonomic continuum (Figure~\ref{fig:mach_cone}b), vehicles cannot translate
sideways. They must steer. The obstacle induces a density gradient, which triggers the
navigational intent ($-\kappa\partial_y\rho_1$). By the constraint, this steering forces
a lateral velocity, and it propagates the lane-changing information at the finite speed $c$.
The traffic advects at a speed $u_0 > c$, so the steering disturbance is swept
backward. It forms a sharp hyperbolic wake, a \emph{Mach cone} of density and orientation,
with half-angle $\alpha = \arcsin(c/u_0)$.

The boundary of this cone carries a critical physical observable that standard models lack.
Across the V-shaped fronts, the steering rate $\dot\theta_1$ peaks. So
the exact nonholonomic reaction \eqref{eq:lambda}, which supplies the centripetal force
$\lambda \approx \rho_0 u_0 (\partial_t\theta_1 + u_0\partial_x\theta_1)$, is largest
along the Mach lines. This maps exactly the regions where the tyres must exert
the highest lateral force. Where the predicted reaction $|\lambda|$ exceeds the available
friction $\mu_s \rho g$, the model flags a localized loss of lateral
adherence (skidding), and it thus links macroscopic traffic variables with vehicle-level mechanics. This is a linear-response indicator, not a quantitative skidding
criterion. An actual loss of adherence next to a near-stationary obstacle is strongly
nonlinear: the local speed $u$ collapses and the small-angle approximation breaks
down. A firm prediction there requires the full nonlinear equations.

\section{Relation to inertial flocking}
\label{sec:flocking}

The microinertia $\chi$, which distinguishes the present theory from standard
active nematics, has a direct counterpart in the physics of bird
flocks~\cite{Attanasi2014,Cavagna2010}.

Field experiments on starling flocks show that collective turns propagate linearly
across the flock, with negligible attenuation. The dispersion is underdamped,
$\omega\simeq\pm c\,q$, rather than diffusive~\cite{Attanasi2014}. This is captured
by the inertial spin model~\cite{Cavagna2015ISM}, in which each bird carries an
orientation and a conjugate spin with its own inertia, so that the orientational
dynamics is second order in time. The present continuum has exactly this structure:
the substructural balance~\eqref{eq:macro-c-b}, with the Frank microstress
$\xii=K\grad\theta$, is second order in time, with $\chi$ the ``behavioural
inertia'', the balance itself the continuum spin dynamics, $\xii=K\grad\theta$ the
bend elasticity of the flock, and
the hyperbolic waves of Section~\ref{sec:benchmark} the turning waves. In this
sense the model is a Capriz-type realization of the inertial spin model.

The closest existing continuum theory is the hydrodynamics of turning
flocks~\cite{YangMarchetti2015}, obtained by coarse-graining the inertial spin
model~\cite{Cavagna2015ISM}. It already couples orientational inertia
to bend elasticity, and it yields anisotropic spin waves. Orientational inertia and
director elasticity are therefore not, in themselves, the novelty here. The novelty
is what the skate constraint adds on top of them.

\emph{(a) Banking force.} The inertial spin model fixes the speed by hand, whereas
here it follows from $\vv=u\nn$, with reaction $\lambda_c=\rho\,u\,\dt\theta$. Since
$\rho\,u\,\dt\theta$ is the centripetal force of a turn, and birds generate it by
banking, the reaction $\lambda_c$ predicts the transverse force field, that is, the
bank angle and the load factor, during collective turns. This is a mechanical
quantity that spin models do not describe.

\emph{(b) Turn--density coupling.} The constraint couples turns to density through
$v_\perp=u_0\theta$ (Section~\ref{sec:benchmark}), so that turning waves are
accompanied by density waves. This is a testable prediction.

\emph{(c) Stress tensor.} The momentum balance with $\TT$ supplies principled
boundary conditions for confined flocks, for obstacles, and for the response to a
predator, which agent-based spin models do not provide.

\emph{(d) Criticality.} The angle $\theta$ is the Goldstone mode of the broken
rotational symmetry, and $\xii=K\grad\theta$ is its stiffness. The scale-free
correlations observed in flocks~\cite{Cavagna2010} arise as its soft modes, and the
control parameter, namely the sign of $\kappa$ in Eq.~\eqref{eq:bench-dir},
tunes the proximity to the banding instability where the correlations diverge.

Topological rather than metric interactions~\cite{Ballerini2008} are a microscopic
feature, absorbed into $K$ and the constitutive laws, and the model is agnostic to
them. A quantitative account of the scale-free correlations requires a fluctuating
hydrodynamics near criticality, for which the present deterministic theory provides
the field-theoretic skeleton. The model is thus best suited to
phenomena where the mechanics matters, such as turns with banking, the coupling
between a manoeuvre and the density, and flocks that interact with boundaries.

\section{Concluding remarks}
\label{sec:outlook}

We have developed a continuum theory for a fluid of elongated particles that
advance along their own axis, subject to a skate constraint. We treated the
constraint as an ideal nonholonomic internal constraint within a continuum with
vectorial microstructure, and we derived its consequences by the principle of virtual power. The central
result is the structure of the reaction. A holonomic constraint on the velocity
gradient produces a reactive stress, but the skate constraint, which is algebraic
in the velocity, produces a reactive body force $\lambda\,\eperp$ that is
transverse to the axis and carries the centripetal term $\rho u\dot\theta$. We
showed that this reaction is workless, so the theory is consistent with the second
law, and we found that the Cauchy stress is non-symmetric, with its skew part
carried by the orientational self-force. We then proved that the planar
Ericksen--Leslie theory, once the constraint is imposed, is a constitutive closure
of the model, and we gave the map between the two descriptions.

The theory unifies, and sheds light on, several collective phenomena. A benchmark
problem showed that the constraint changes the very type of the governing
equations, turning the parabolic orientational diffusion of a passive nematic into
hyperbolic orientation--density waves. In vehicular traffic, the one-dimensional
reduction recovers classical macroscopic models, while the two-dimensional extension
describes steering, the lateral tyre force, and the road geometry. In the physics
of bird flocks, the microinertia identifies the model as a continuum realization of
the inertial spin model, adding to it a banking force and a coupling between turns
and density.

The model also provides a rigorous mechanical foundation for the
phenomenological Toner--Tu equations of polar active fluids. It reproduces their
hyperbolic waves, but it adds two fundamental ingredients. First, it fixes the
empirical convective coefficient $\lambda_1$ constitutively, tying it to the
flow-alignment of the units ($\gamma_2/\gamma_1$). Second, it introduces the
transverse reaction $\lambda$, which is completely absent in unconstrained symmetric
stresses, supplying the explicit centripetal force that the manoeuvres demand.

Several developments are natural. The most structural is the generalization to three
dimensions, where the orientation is a director on the sphere $\Sph^2$ and the
reaction is a body force in the normal plane. This is the natural setting for starling
murmurations and drone swarms. A rigorous overdamped limit $\chi\to0$ would connect
the theory to the self-organized hydrodynamics, while the full stress tensor it
provides invites principled boundary conditions for confined flocks and for the
response to obstacles or a predator. Closer to the collective-motion literature, a
fluctuating hydrodynamics built on the present deterministic skeleton would address
the scale-free correlations observed near the banding instability.

The change of type from parabolic to hyperbolic dynamics has a concrete payoff, which
the benchmark of Section~\ref{sec:traffic-numerical} already illustrates. A
two-dimensional flow past an obstacle develops a hyperbolic Mach cone of density and
orientation, in place of the diffusive boundary layer of the parabolic models. The
reaction field $\lambda(x,y,t)=\rho\,u\,\dt\theta$ is the natural observable to
exploit further. Where its magnitude exceeds a structural or frictional threshold, the
theory flags a macroscopic breakdown: loss of tyre adherence (skidding) in
traffic, or the onset of jamming and arching (clogging) when a flock is forced
through a narrow bottleneck. This is a prediction that quantitative simulation and
experiment can now test.

\clb{More broadly, the same triple $(u,\nn,\lambda)$ --- an advancing speed, a body
axis, and the transverse reaction that enforces the no-side-slip condition --- recurs
across many systems of oriented self-propelled agents, which the theory places on a
common footing (Table~\ref{tab:dictionary}). The rigid form of the constraint is best
realized where lateral slip is mechanically forbidden: wheeled vehicles, fixed-wing
drones and unicycle-type ground robots in swarm robotics, whose kinematic model is
precisely the Chaplygin sleigh, and gliding filaments or bacteria confined in
microfluidic channels. This also sharpens the distinction, drawn in
Section~\ref{sec:traffic}, between holonomic pedestrian crowds, which can step
sideways, and nonholonomic vehicular flow. For free-swimming microswimmers and sperm
the constraint is softer --- an anisotropic drag rather than a rigid prohibition, as
discussed in Section~\ref{sec:sota} --- so the theory applies there only in this
frictional, non-ideal sense. Three further directions seem natural: active matter on
curved substrates and shells, where the director geometry of
Section~\ref{sec:traffic} becomes intrinsic (a setting the author has studied with
Turzi); the interaction of these flows with obstacles, of
which the Mach cone is the simplest instance; and active mechanical metamaterials
whose oriented units carry an internal no-slip reaction. We offer these as
conjectures. The appeal of the mechanical viewpoint is that in each case the reaction
$\lambda$ is a definite, measurable force rather than a fitting parameter.}

\begin{table}[t]
\centering
{\footnotesize
\begin{tabular}{@{}llll@{}}
\toprule
System & Speed $u$ & Axis $\nn$ & Reaction $\lambda$ \\
\midrule
Traffic & vehicle speed & heading & lateral tyre force \\
Flock & flight speed & body axis & banking / centripetal force \\
Drone swarm & forward speed & heading & lateral manoeuvring force \\
Robot swarm & forward velocity & body orientation & steering force \\
Microswimmers & swimming speed & polarity & transverse hydrodynamic demand \\
Active metamaterial & propagation speed & local director & internal constraint reaction \\
\bottomrule
\end{tabular}}
\caption{\clb{The triple $(u,\nn,\lambda)$ across candidate realizations of the skate
constraint: an advancing speed $u$, a body axis $\nn$, and the transverse reaction
$\lambda$ that enforces $\vv=u\nn$. The upper rows are rigid realizations; the lower
rows are softer or more speculative.}}
\label{tab:dictionary}
\end{table}



\end{document}